\documentclass[12pt]{article}
\usepackage{amsmath}
\usepackage{amsfonts}
\usepackage[style=authoryear, natbib=true, url=false, isbn=false, doi=false, backend=bibtex]{biblatex}
\usepackage{datetime}
\usepackage[margin=0.75in]{geometry}
\usepackage{graphicx}
\usepackage[doublespacing]{setspace}
\usepackage{subfigure}
\usepackage{multirow}
\usepackage{algorithm}
\usepackage[noend]{algpseudocode}
\newdateformat{mydate}{\THEDAY\ \monthname[\THEMONTH]\ \THEYEAR}

\bibliography{/Users/dturek/Dropbox/Berkeley/References/ZoteroLibrary}
\renewbibmacro{in:}{}

\newcommand{\blind}{0}

\begin{document}

\if0\blind {
\title{Automated Parameter Blocking for Efficient Markov-Chain Monte Carlo Sampling}
\author{Daniel Turek$^{*1,2}$, Perry de Valpine$^{2}$, \\ Christopher J. Paciorek$^{1}$, Clifford Anderson-Bergman$^{1,2}$}
\date{\mydate\today}
\maketitle
\vspace{0.25in}
\begin{center}
$^*$Corresponding Author \\
dturek@berkeley.edu \\
\ \\
$^1$Department of Statistics \\
University of California, Berkeley \\
\ \\
$^2$Department of Environmental Science, Policy, and Management \\
University of California, Berkeley \\
\end{center}
\thispagestyle{empty}
\newpage
} \fi

\if1\blind {
\ 
\vspace{0.5in}
\begin{center}
\LARGE Automated Parameter Blocking for Efficient Markov-Chain Monte Carlo Sampling
\end{center}
\vspace{0.5in}
} \fi

\thispagestyle{empty}

\begin{abstract}
Markov chain Monte Carlo (MCMC) sampling is an important and commonly used tool for the analysis of hierarchical models.  Nevertheless, practitioners generally have two options for MCMC: utilize existing software that generates a black-box ``one size fits all" algorithm, or the challenging (and time consuming) task of implementing a problem-specific MCMC algorithm.  Either choice may result in inefficient sampling, and hence researchers have become accustomed to MCMC runtimes on the order of days (or longer) for large models.  We propose an automated procedure to determine an efficient MCMC algorithm for a given model and computing platform.  Our procedure dynamically determines blocks of parameters for joint sampling that result in efficient sampling of the entire model.  We test this procedure using a diverse suite of example models, and observe non-trivial improvements in MCMC efficiency for many models.  Our procedure is the first attempt at such, and may be generalized to a broader space of MCMC algorithms.  Our results suggest that substantive improvements in MCMC efficiency may be practically realized using our automated blocking procedure, or variants thereof, which warrants additional study and application.
\end{abstract}

\vspace{0.5in}

\textbf{Keywords:} \\
\indent \indent MCMC, Metropolis-Hastings, Block sampling, Integrated autocorrelation time, Mixing.

\thispagestyle{empty}
\newpage

\section{Introduction} \label{sec:introduction}

Markov chain Monte Carlo (MCMC) has become a core computational method for many statistical analyses.  These include routine Bayesian analyses, but also hybrid algorithms that use MCMC as one component, such as Monte Carlo Expectation Maximization \citep[MCEM;][]{Caffo2005} or data cloning \citep{Lele2007}.  Nevertheless, the automated generation of black-box MCMC algorithms, as occurs in generally available software, does not necessarily result in efficient MCMC sampling.  Analysts are thereby accustomed to MCMC run times measured in minutes, hours or even days for large hierarchical models.  Computation time is frequently the limiting factor, either limiting the range of models considered, or limiting the potential for performing diagnostics and comparisons using methods such as bootstrapping \citep{Efron1994}, cross validation \citep{Gneiting2007}, or calibration of posterior predictive p-values \citep{Hjort2006}, among others.  Therefore, any widely applicable improvements to MCMC performance may greatly improve the practical analyses of large hierarchical models.

Among the many MCMC sampling algorithms developed to improve MCMC efficiency, one of the most basic approaches has been block sampling: jointly updating multiple dimensions of a target distribution simultaneously \citep{Roberts1997, Sargent2000}.  When one or more dimensions of the posterior distribution are correlated, joint sampling of these dimensions (with any variety of block samplers) can increase sampling performance relative to updating each dimension independently \citep[\emph{e.g.,}][]{Liu1994}.  Despite wide recognition of the usefulness of this basic idea for designing efficient MCMC algorithms, there has been no automated method for choosing blocks to optimize -- or at least greatly improve -- performance.  Here we develop such a method.

Existing theoretical work comparing block samplers to univariate samplers \citep[among others]{Mengersen1996, Roberts1996a, Roberts1997a} has provided many insights but falls short of providing a complete assessment of MCMC efficiency for several reasons.  First, it uses MCMC convergence rates as the metric for comparison, without consideration of the computational demands of block sampling.  Instead, our viewpoint is that any measure of MCMC efficiency must incorporate both the convergence rate and the computational requirements of achieving improvements in convergence rate.  This may give a different picture of the actual efficiency of a sampling algorithm.  Accelerated convergence at an extreme computational cost is obviously not optimal.  Second, the computational requirements of different steps will vary greatly across platforms, depending on such factors as processor, memory architecture, use of efficient linear algebra packages, etc.  Therefore, even if theoretical comparisons were extended to incorporate aspects of computation, the best block sampling scheme would remain platform-dependent.  It is important to recognize that computational costs affect not only the proposal step -- such as the cost of generating a multivariate normal proposal -- but also model computations and density evaluations.  Some parts of hierarchical models may inherently involve expensive computations, which can impact the relative efficiency of different blocking schemes.  Third, existing theories and methods presume that some wise, manual selection of blocks may be feasible, based for example on an understanding of the model structure, which leads to understanding which posterior dimensions may be correlated.  In general, however, it is difficult to know \emph{a priori} which dimensions will be correlated, which is one purpose of automating a procedure like MCMC in the first place.

Here, we present a procedure for the automated exploration of MCMC blocking schemes, seeking a highly efficient MCMC algorithm specific to the hierarchical model and computing environment at hand.  This represents a higher level of automated algorithm generation than is provided by existing software, which serve to produce ``one size fits all" MCMC algorithms.  The family of BUGS packages \citep[WinBUGS, JAGS, and OpenBUGS;][]{Lunn2000, Plummer2011, Lunn2012} assigns samplers based on local characteristics of each model parameter, using a combination of Gibbs sampling, adaptive rejection sampling, slice sampling, and, in limited cases, block sampling.  Other MCMC packages including ADMB \citep{Skaug2006} and Stan \citep{StanDevelopmentTeam2014} use Hamiltonian MCMC sampling \citep{Neal2011}, which may generally be more efficient but nevertheless represents a static approach to MCMC algorithm generation.  Yet other promising methods such as Langevin sampling \citep{Marshall2012} are not incorporated into software commonly used by practitioners.  For simplicity, we restrict our attention to univariate and blocked adaptive random walk sampling.  However, the main concept of exploring the space of parameter blocks to improve MCMC efficiency generalizes to allow the use of other sampling methods.

In section \ref{sec:efficiency}, we examine the pros and cons of univariate versus block random walk sampling, both in terms of algorithmic and computational efficiencies.  From these considerations we conclude that the combination of samplers that yields optimal MCMC efficiency (defined as an MCMC algorithm's rate of generating effectively independent samples) will be model- and platform-specific.  Section \ref{sec:autoblock} introduces a procedure for automated blocking of hierarchical model parameters, designed to maximize the resulting MCMC algorithm efficiency.  The main idea of this procedure is to iteratively cluster model parameters based upon empirical posterior correlations, then intelligently subdivide the hierarchical clustering tree (a dendrogram) to determine blockings of parameters that result in efficient MCMC sampling.  A series of simulated and real data examples given in section \ref{sec:performance} demonstrate that this procedure can improve MCMC efficiency many-fold over statically defined MCMC algorithms and can dynamically generate algorithms comparable in performance to model-specific, hand-tuned algorithms.  We close with a discussion in section \ref{sec:discussion}.

\section{MCMC Algorithms: Definitions and Efficiency} \label{sec:efficiency}

In this section, we first define a space of valid MCMC algorithms.  Then, we examine two dominant contributors to the efficiency of an MCMC algorithm: the algorithmic capability to produce effectively-independent samples from the target distribution, and the computational demands of the algorithm in generating MCMC samples; these are composed to form our metric of overall MCMC efficiency.  Drawing upon existing asymptotic theory of MCMC sampling, the scaling of computational costs of different sampling schemes, and on several illustrative examples, we argue that achieving an optimally efficient MCMC algorithm for a specific model by pure theory is prohibitively difficult.  That conclusion motivates our approach of computationally optimizing -- or at least greatly improving -- MCMC performance through automated exploration of a space of valid MCMC algorithms.

\subsection{MCMC Definition} \label{mcmcdef}

We assume a given, fixed, hierarchical model $\mathcal{M}$, which may be represented as a directed acyclic graph where vertices represent top-level model parameters, latent states, or fixed observations (data), and edges represent dependencies between these components.  Denote the set of all top-level model parameters and latent states (the unknown components for which we may seek inferences) as $\Theta = \{ \theta_1, \ldots, \theta_d \}$, which will be referred to as the parameters of $\mathcal{M}$.

An MCMC algorithm may be defined in terms of its sampling scheme over $\Theta$.  Let $b$ be any non-empty subset (``block") of $\Theta$, and $u \in U$ be any valid MCMC sampling (or ``updating") method such as slice sampling or conjugate Gibbs sampling \citep[see][for a broad overview of MCMC sampling methods]{Gilks2005}.  We define a valid MCMC sampler $\psi = u(b)$ as the application of $u$ to $b$, where $b$ satisfies any assumptions of $u$ (\emph{e.g.,} conjugacy).  In addition to satisfying standard properties of Markov chains \citep[see, for example,][]{Robert2004}, we define a valid MCMC algorithm as any set of samplers $\Psi = \{ \psi_1, \ldots, \psi_k \}$, where $\psi_i = u_i(b_i)$ for $i = 1, \ldots, k$, satisfying $\cup_{i=1}^k b_i = \Theta$; that is, the MCMC algorithm updates each model parameter in at least one sampler.  We represent the chain of samples generated from successive applications of $\Psi$ as $X^{(0)}, X^{(1)}, \ldots$, where sample $X^{(i)}$ implies model state $\Theta = X^{(i)}$, $X^{(0)}$ is the set of initial values, $X^{(i)} = (X^{(i)}_1, \ldots, X^{(i)}_d)$, and $X_k = \{ X^{(0)}_k, X^{(1)}_k, \ldots \}$ is the scalar chain of samples of $\theta_k$ (for $k = 1, \ldots, d$).

This paper focuses attention on the restricted set of sampling methods $U_0 = \{ u_\text{scalar}, u_\text{block} \}$, where $u_\text{scalar}$ denotes univariate adaptive random walk Metropolis-Hastings sampling \citep[hereafter, scalar sampling;][]{Metropolis1953, Hastings1970}, and $u_\text{block}$ denotes the multivariate generalization of this algorithm \citep[hereafter, block sampling;][]{Haario1999}, with the practical restriction that any $\psi = u_\text{block}(b)$ satisfies $| b | > 1$.  The $u_\text{scalar}$ algorithm adaptively tunes the proposal scale, while $u_\text{block}$ additionally tunes the proposal covariance \citep{Roberts2009}.  All scalar and block samplers asymptotically achieve the theoretically optimal scaling of proposal distributions (and therefore acceptance rates, and mixing) as derived in \citet{Roberts1997a}, and implement adaptation routines as set out in \citet{Shaby2011}.

For hierarchical model $\mathcal{M}$ with parameters $\Theta$, our studies focus almost exclusively on the set of MCMC algorithms $\boldsymbol{\Psi}_\mathcal{M}$, which contains all algorithms of the form $\Psi = \{ \psi_1, \ldots, \psi_k \}$, where $\psi_i = u_i(b_i)$, each $u_i \in U_0$, and the sets $b_i$ form a partition of $\Theta$.  We specifically name two algorithms in $\boldsymbol{\Psi}_\mathcal{M}$ which are boundary cases.  The first consists of $d$ scalar samplers: $\Psi_\text{scalar} = \{ \psi_1, \ldots, \psi_d \}$, where each $\psi_i = u_\text{scalar}(\theta_i)$.  The second has a single block sampler for all parameters: $\Psi_\text{block} = \{ u_\text{block}(\Theta) \}$.  Implicit is the assumption that each $\theta_i$ is continuous-valued, which is the case throughout this paper.

\subsection{Algorithmic Efficiency} \label{sec:algorithmiceff}

We first consider MCMC algorithmic efficiency, independent of any computational requirements.  This measure of efficiency solely represents the best mixing, or equivalently the least autocorrelation, or the highest effective sample size, without consideration for the computational (time) requirements of generating a set of samples.  After reviewing the definition of MCMC algorithmic efficiency which is based upon integrated autocorrelation time, we study the use of $\Psi_\text{scalar}$ or $\Psi_\text{block}$ for particular choices of $\mathcal{M}$, and quantify the effects on this measure of efficiency.

As in \citet{Roberts2001}, we define MCMC algorithmic efficiency as the effective sample size divided by the chain length.  This represents the rate of production of effectively independent samples per MCMC sample.  The effective sample size (ESS) of an MCMC chain is defined as $\text{ESS} = N / \tau$, where $N$ is the chain length and $\tau$ is the integrated autocorrelation time.  For a scalar chain of samples $X_0, X_1, \ldots$, which is assumed to have converged to its stationary distribution, \citet{Straatsma1986} define the integrated autocorrelation time as $\tau = 1 + 2 \sum_{i=1}^{\infty} \text{cor} (X_0, X_i)$.  $\tau$ may be interpreted as the number of MCMC samples required, on average, for an independent sample to be drawn.  Our measure of algorithmic efficiency is thus $\tau^{-1}$, the number of effective samples per actual sample \citep{Thompson2010}.  $\tau^{-1}$ also characterizes the speed at which expectations of arbitrary functions of the sample values approach their stationary values \citep{Roberts1997}, and no less satisfies the natural intuition that larger values indicate better performance.

For MCMC algorithm $\Psi$ acting on model $\mathcal{M}$ with parameters $\Theta$, we define the algorithmic efficiency of each $\theta \in \Theta$ as $A(\Psi, \theta) = \tau^{-1}$, where $\tau$ is the integrated autocorrelation time of the samples of $\theta$ generated from repeated application of $\Psi$.  Overloading notation, we define the algorithmic efficiency of MCMC algorithm $\Psi$ as $A(\Psi) = \text{min}_{\theta \in \Theta} \, A(\Psi, \theta)$.  This definition is motivated by noting that often an MCMC produces seemingly good mixing of many model dimensions but poor mixing of just a few dimensions.  In this case, the poorly mixing dimensions will limit the validity of the entire posterior sample (although this is not universally true of all model structures).  Therefore, we take the conservative approach, and our general aim is to maximize the algorithmic efficiency for the parameter exhibiting the slowest mixing.

We now study the potential for algorithmic \emph{inefficiency} that may result from scalar or block sampling, by examining situations in which each are particularly ill-suited.

\subsubsection*{Efficiency loss from block sampling}

Consider MCMC algorithm $\Psi_\text{block} \in \boldsymbol{\Psi}_\mathcal{M}$.  Application of $\Psi_\text{block}$ generates a random proposal vector $X^* \in \mathbb{R}^d$, then jointly accepts or rejects $X^*$.  In the framework of the sampling method $u_\text{block}$, $X^* \sim \text{N}_d(\mu, \sigma^2_d \Sigma)$, where $\mu$ and $\Sigma$ vary according to current state of the MCMC chain and properties of the target stationary distribution, but not proportionally to $d$.  \citet{Roberts1997a} show that in order to achieve the asymptotically optimal acceptance rate, and therefore sample chain mixing, $\sigma^2_d \propto d^{-1}$.  As a consequence of this attenuation in the proposal variance, the rate at which $\Psi_\text{block}$ explores the space of $\mathbb{R}^d$, and accordingly $A(\Psi_\text{block})$, is proportional to $d^{-1}$.  This result applies equivalently to block samplers $\psi = u_\text{block}(b)$ acting on subsets $b \subset \Theta$, where the algorithmic efficiency (for the elements of $b$) achieved by application of $\psi$ is inversely proportional to the number of elements in $b$.

This result has an important implication on block sampling.  All other factors being equal (\emph{e.g.}, effect of posterior correlations), a block sampler of dimension $k$ must generate $k$ times more samples to have the same effective sample size as those samples produced through scalar sampling \citep{Roberts2001}.  This inefficiency is inherent to block sampling and scales with the dimension of any block sampler.

\subsubsection*{Efficiency loss from scalar sampling}

To study the potential loss of algorithmic efficiency which may result from scalar sampling under $\Psi_\text{scalar} \in \boldsymbol{\Psi}_\mathcal{M}$, we consider correlated posterior distributions.  It is well-understood that strong posterior correlations can retard the speed of convergence of MCMC sampling \citep[\emph{e.g.,}][]{Roberts1997, Gilks2005}, and that block sampling can alleviate this.  However, the nature of this inefficiency is not characterized, in particular as a function of the degree of correlation and number of dimensions exhibiting correlation.  We undertake a simulation study, to gauge how these factors effect algorithmic efficiency.  Consider target distribution $\text{N}_d(0, \Sigma)$, where the covariance (equivalently, correlation) matrix $\Sigma$ consists of 1s on the main diagonal and $\rho$ elsewhere, $|\rho| < 1$.  We consider the algorithmic efficiencies of individual model parameters under scalar sampling, $A(\Psi_\text{scalar}, \theta)$ for $\theta \in \Theta$.

\begin{figure}[h]
\centerline{\includegraphics[scale=1.0]{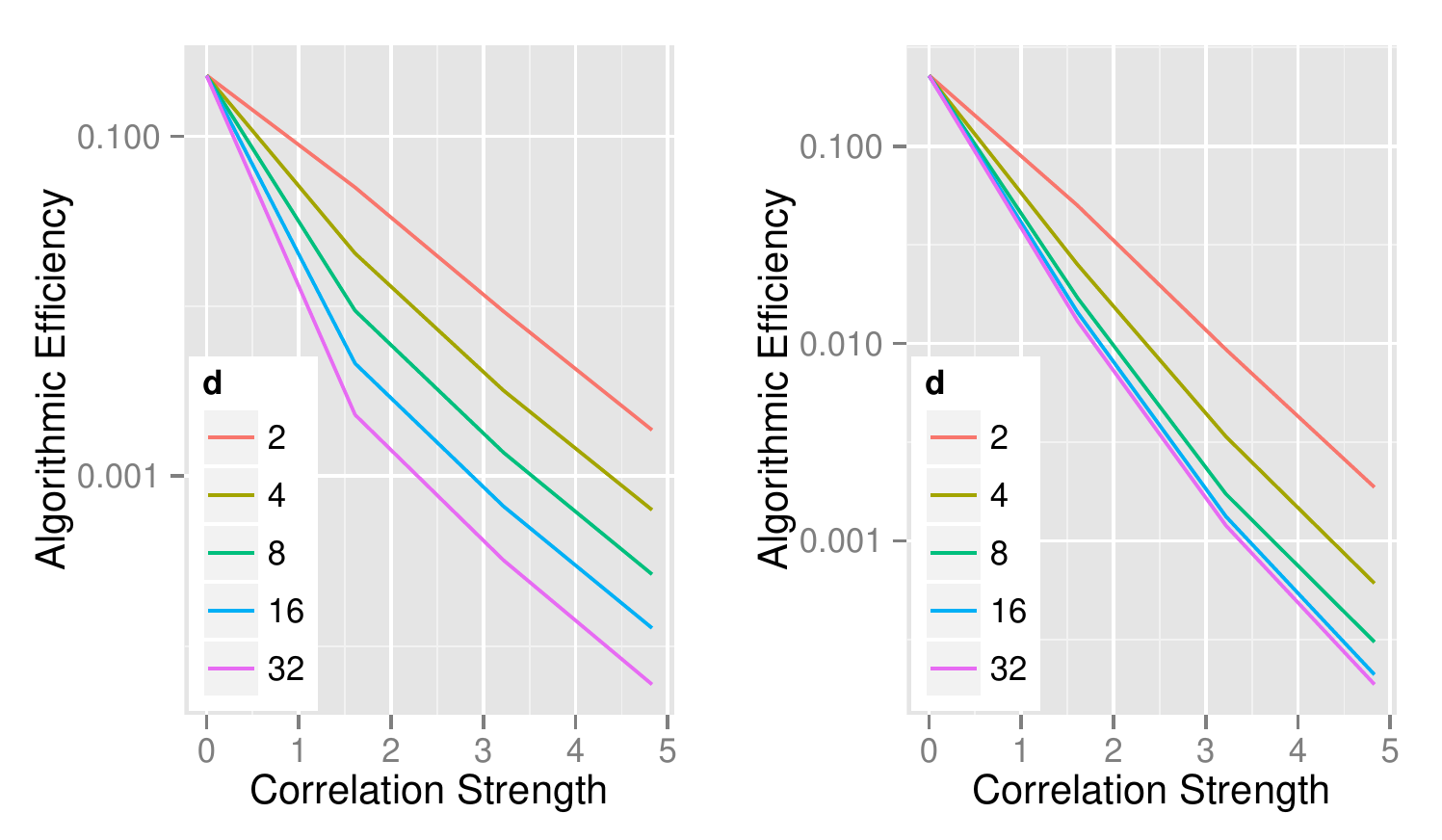}}
\caption{MCMC algorithmic efficiencies for different values of model dimension (d), and intra-group correlation ($\rho$).  The quantity $-\log(1-\rho)$ is plotted on the horizontal axis.  Model structures include constant off-diagonal elements (equal to $\rho$) in the induced correlation matrix (left), and exponentially decaying correlations (right).}
\label{fig:algorithmicEfficiency}
\end{figure}

Empirically, we observe that each $A(\Psi_\text{scalar}, \theta)$ tends toward zero as $\rho$ approaches one, or as $d$ diverges ($\rho \neq 0$).  The nature of these relationships is characterized on a log-log scale in Figure \ref{fig:algorithmicEfficiency} (left), where the horizontal axis plots $-\log(1-\rho)$, such that positive horizontal shifts represent $\rho$ exponentially approaching the boundary $\rho=1$, or perfect correlation between parameters.  As one would expect, when $\rho=0$ all values of $d$ yield identical algorithmic efficiency.  However, when $\rho > 0$ we enter a linear regime where each $A(\Psi_\text{scalar}, \theta)$ exponentially decays towards zero.  Even for fixed $d$, algorithmic efficiency under $\Psi_\text{scalar}$ can be arbitrarily poor as $\rho$ approaches unity.

It may be extreme to assume a target distribution with arbitrarily large blocks of parameters that exhibit arbitrarily high pairwise-correlation.  As an alternative, we consider the same multivariate normal form, but with elements of $\Sigma$ given as $\sigma_{i,j} = \rho^{|i-j|}$, $|\rho|<1$, as might occur in the covariance structure of a spatial model \citep[p.27]{Banerjee2003}.  The algorithmic efficiency of the first model parameter -- since elements of $\Theta$ are no longer exchangeable -- is shown in Figure \ref{fig:algorithmicEfficiency} (right), where it displays similar attenuation as in the prior example.  Most notably, we now observe the incremental effect of $d$ diminishing as $d$ increases, consistent with the covariance structure.

Through a combination of theory and simulated examples, we observe that the algorithmic efficiency achieved under $\Psi_\text{scalar}$ or $\Psi_\text{block}$ will depend non-trivially upon the model dimension, and the extent and structure of the posterior correlation.  We have only considered two simple, highly modular and systematic posterior correlation structures.  In practice, any model of interest will demonstrate a substantially more complex correlation structure (which is unknown, anyway), making a full analytical study of MCMC algorithmic efficiency difficult.  No less, we have only considered the two boundary-case algorithms $\Psi_\text{scalar}$ and $\Psi_\text{block}$ in $\boldsymbol{\Psi}_\mathcal{M}$.  Our desire to derive general results for algorithmic efficiency will be complicated substantially further when we consider the complete set $\boldsymbol{\Psi}_\mathcal{M}$.
\subsection{Computational Efficiency} \label{sec:computation}

While section \ref{sec:algorithmiceff} considered the efficiency of MCMC algorithm $\Psi$ at producing independent samples without regard for computation time, we now consider the computational requirements of $\Psi$, measured in units of algorithm runtime per MCMC iteration.  We focus on computations for model density evaluations.  There are also book-keeping and loop-iteration costs, which we label as algorithm overhead.  These overhead terms comprise a small fraction of total computation, and we safely disregard them.  Denote the computational requirement of $\Psi$ as $C(\Psi)$, and again overload notation to define $C(\psi)$ as the computational requirement of a single sampler $\psi$.  For $\Psi = \{ \psi_1, \ldots, \psi_k\}$, $C(\Psi) = \sum_{i=1}^k C(\psi_i)$.  As far as we are aware, an analysis of MCMC efficiency which incorporates $C(\Psi)$ has not been carried out to date.  Literature which does address MCMC efficiency typically recognizes that a computational aspect exists, but then focuses solely on $A(\Psi)$, \emph{e.g.}, \citet{Roberts1997}.

We now present an accounting of the main contributions to $C(\Psi_\text{scalar})$ and $C(\Psi_\text{block})$, general for any $\mathcal{M}$.  To support our accounting, we denote the set of all fixed and known components of $\mathcal{M}$ (\emph{e.g.,} observations, other data) as $Y$, which is disjoint from the unknown set of parameters $\Theta = \{ \theta_1, \ldots, \theta_d\}$.  For each $\theta_i \in \Theta$, let $x_i \subset \Theta \cup Y$ be the set of model components which immediately depend on $\theta_i$, or the set of direct descendants of $\theta_i$ in the model graph introduced in section \ref{mcmcdef}.  Finally, we denote the computational cost of evaluating the density functions corresponding to any subset $x \subset \Theta \cup Y$ as $\text{dens}(x)$.

\subsubsection*{Scalar Sampling Computation}

On each iteration of $\Psi_\text{scalar} = \{ \psi_1, \ldots, \psi_d \}$, each sampler $\psi_i$ will incur computational cost $C(\psi_i) = \text{dens}(\theta_i) + \text{dens}(x_i) + O(1)$.  The trailing constant term represents generation of the proposal value and making the MH decision (generation from normal and uniform distributions, respectively).  The total computational requirement of $\Psi_\text{scalar}$ is thus
$$C(\Psi_\text{scalar}) = \sum_{i=1}^d C(\psi_i) = \sum_{i=1}^d \text{dens}(\theta_i) + \sum_{i=1}^d \text{dens}(x_i) + O(d).$$
Note that under $\Psi_\text{scalar}$, each density evaluation $\text{dens}(\theta_i)$ must occur independently.  This is true even when the evaluation of a particular $\text{dens}(\theta_i)$ term necessitates the calculation of a subsuming multivariate density -- in the most extreme case, $\text{dens}(\Theta)$.  Thus, in the worst case, a single MCMC iteration of $\Psi_\text{scalar}$ could incur $d$ evaluations of the entire joint density of $\Theta$.  A similar computational explosion can result from the calculation of each $\text{dens}(x_i)$ term.

\subsubsection*{Block Sampling Computation}

We now consider the components of $C(\Psi_\text{block})$.  On each iteration of $\Psi_\text{block}$, the sole sampler $u_\text{block}(\Theta)$ requires evaluation of the complete prior and likelihood densities, $\text{dens}(\Theta \cup Y)$.  This is notably different from the density evaluation terms appearing in $C(\Psi_\text{scalar})$, in that it incurs only a \emph{single} evaluation of the complete joint model density.  In addition, the adaptation routine of $u_\text{block}(\Theta)$ requires a Cholesky factorization of the adapted covariance matrix, which requires $O(d^3)$ operations to calculate in full generality \citep[p.176]{Trefethen1997}.  This factorization occurs every AI iterations, where AI is the adaptation interval of $u_\text{block}$, and therefore has an amortized computational cost of $O(d^3 / \text{AI})$.  Each iteration of $u_\text{block}$ requires generating a $d$-dimensional multivariate normal proposal, which requires $O(d^2)$ operations, and performing a single constant-time MH decision, which is dropped as a lower-order term.  The total amortized computational requirement of $\Psi_\text{block}$ may be written as
$$C(\Psi_\text{block}) = \text{dens}(\Theta \cup Y) + O(d^3 / \text{AI}) + O(d^2).$$

\subsubsection*{Timing Comparison}

We have seen that $C(\Psi_\text{block})$ is at least quadratic in $d$, and technically cubic but with a small leading coefficient.  Depending on the distributional structure of $\Theta$, the density evaluations comprising $C(\Psi_\text{scalar})$ may be unwieldy.  The relative magnitude of these competing terms is difficult to intuitively gauge, so to gain practical insight, we perform a timing study of the $\Psi_\text{scalar}$ (All Scalar) and $\Psi_\text{block}$ (All Blocked) algorithms.  Three models involving no likelihood components are considered, with prior structures on $\Theta$ given as:
\begin{itemize}
\item $\theta_i \sim \text{N}(\mu, \sigma)$ for each $\theta_i \in \Theta$
\item $\theta_i \sim \text{Gamma}(\alpha, \beta)$ for each $\theta_i \in \Theta$
\item $\Theta \sim \text{N}_d(\mu, \Sigma)$
\end{itemize}

\begin{figure}[h]
\centerline{\includegraphics[scale=1.0]{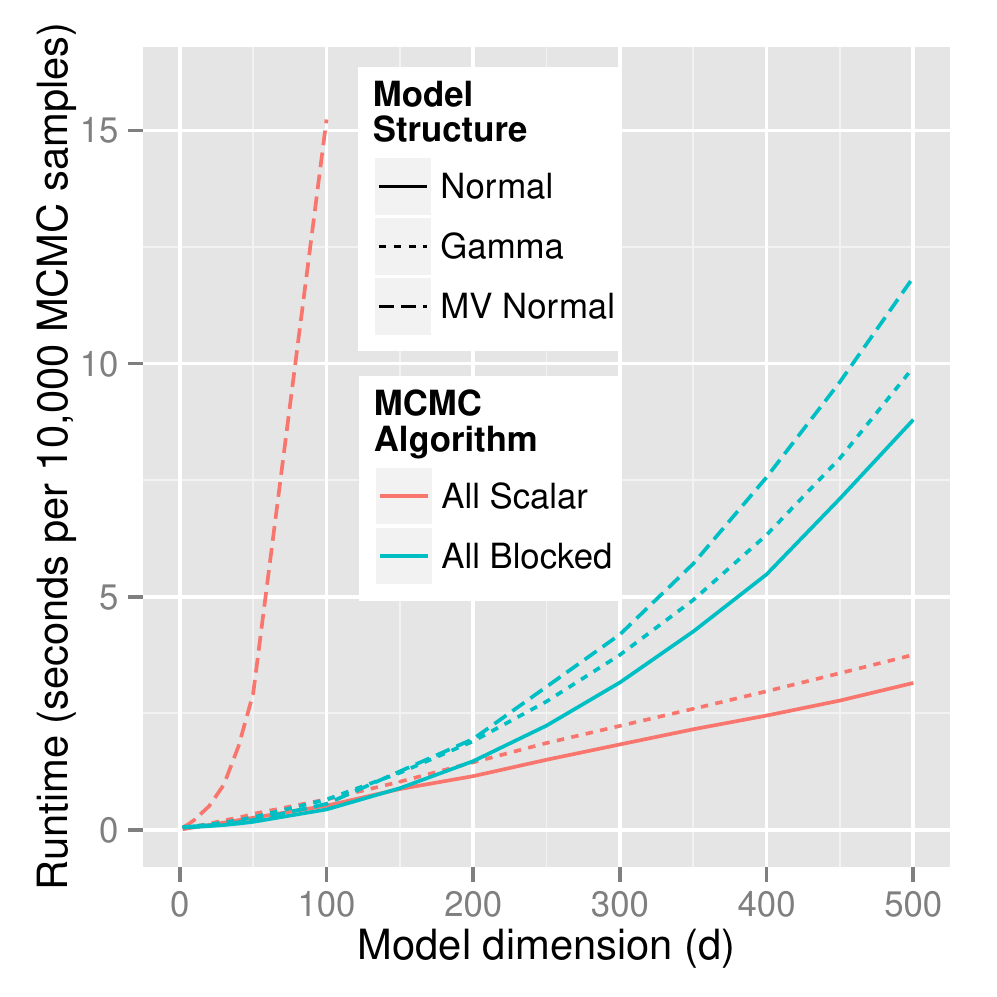}}
\caption{MCMC runtimes for the All Scalar and All Blocked algorithms, for univariate normal, univariate gamma, and multivariate normal model structures.}
\label{fig:computationalRequirement}
\end{figure}

Figure \ref{fig:computationalRequirement} presents timing results measured in seconds per 10,000 iterations, plotted against dimension $d$, without consideration of algorithmic efficiency (section \ref{sec:algorithmiceff}).  There are a number of interesting features, which we briefly summarize.  $C(\Psi_\text{scalar})$ is $O(d)$ when each $\theta_i$ independently follows a univariate distribution, and therefore $\sum_{i=1}^d \text{dens}(\theta_i) = \text{dens}(\Theta) \leq d \cdot K$, where $K = \text{max}_{\theta \in \Theta} \; \text{dens}(\theta)$.  For all practical purposes, it appears that $C(\Psi_\text{block})$ is $O(d^2)$, as the cubic coefficient $1 / \text{AI} = 1 / 200$ is relatively small.  $C(\Psi_\text{block})$ is largely resilient to changes in the underlying distribution of $\Theta$; univariate gamma densities are more costly than normal densities, as we would expect, and as for $C(\Psi_\text{scalar})$; and the multivariate normal structure even slightly more so.  Perhaps most striking, $C(\Psi_\text{scalar})$ is $O(d^3)$ when the underlying distribution of $\Theta$ is multivariate, since each multivariate normal density evaluation is $O(d^2)$, which occurs $d$ times for each iteration of $\Psi_\text{scalar}$.  Although both are cubic in $d$, $C(\Psi_\text{scalar})$ dwarfs $C(\Psi_\text{block})$ due to the difference in the leading cubic coefficients.

\subsection{Overall Efficiency} \label{sec:overallefficiency}

We have examined both the algorithmic and computational efficiency of MCMC algorithms, each of which fundamentally affect overall MCMC efficiency.  We define the overall efficiency of $\Psi$ simply as $E(\Psi) = A(\Psi) / C(\Psi)$.  We consider this to be a sensible measure of the overall efficiency of an MCMC algorithm, since $E(\Psi)$ may be interpreted as the number of effective samples produced per unit of MCMC algorithm runtime, for the slowest mixing model parameter.  If we can construct $\Psi$ to maximize $E(\Psi)$, then $\Psi$ is the most time-efficient MCMC algorithm for generating effectively independent samples to approximate the true, joint posterior distribution of $\Theta$.

That being said, from our examination of algorithmic and computational efficiency, it is not immediately clear how to balance tradeoffs between $A(\Psi)$ and $C(\Psi)$ to maximize $E(\Psi)$.  We have generally considered the two boundary-case algorithms $\Psi_\text{scalar}$ and $\Psi_\text{block}$, but a huge number of intermediate algorithms exist.  For $\Psi \in \boldsymbol{\Psi}_\mathcal{M}$, we may gain useful insights regarding the factors affecting $E(\Psi)$ in terms of the properties of each $\psi_i$.  $C(\Psi) = \sum_{i=1}^k C(\psi_i)$, and the values $A(\Psi, \theta_i)$ which determine $A(\Psi)$ each result from a single application of scalar or block sampling -- although this neglects the phenomenon where improving $A(\Psi, \theta_i)$ may affect $A(\Psi, \theta_j)$, $i \neq j$.  However, finding a global optimum $\Psi_\text{opt} = \text{argmax}_{\Psi \in \boldsymbol{\Psi}_\mathcal{M}} E(\Psi)$ poses a combinatorial challenge.  Instead of seeking $\Psi_\text{opt}$, we now propose an iterative procedure to navigate $\boldsymbol{\Psi}_\mathcal{M}$, with the aim of maximizing $E(\Psi)$ to the degree possible.

\section{Automated Blocking} \label{sec:autoblock}

In this section, we propose an iterative, self-tuning procedure for automated blocking of hierarchical model parameters to produce an efficient MCMC algorithm.  This procedure uses the empirical posterior correlation to cluster groups of correlated parameters into sampling blocks.  A hierarchical clustering tree of model parameters is constructed, and subsequently cut at some height (selected using a finite search) to produce parameter groups each exhibiting a minimal intra-group posterior correlation.  This procedure is iterated, so that as MCMC efficiency improves, the empirical posterior correlations are more accurate, and the resulting tree and parameter groups stabilize.  The end-result is a partition of the model parameters, which uniquely specifies an MCMC algorithm $\Psi \in \boldsymbol{\Psi}_\mathcal{M}$ employing scalar and block sampling, for which the overall efficiency $E(\Psi)$ (section \ref{sec:overallefficiency}) is increased to the degree possible.  We also discuss more sophisticated approaches, but our heuristic approach allows huge efficiency gains in some cases and establishes the basic procedure.

\subsection{Procedure}

Assume a given, fixed, hierarchical model $\mathcal{M}$, with parameters $\Theta = \{ \theta_1, \ldots, \theta_d \}$.  Algorithm \ref{alg:autoblock} presents pseudocode for our automated blocking procedure, which produces MCMC algorithm $\Psi_\text{AutoBlock} \in \boldsymbol{\Psi}_\mathcal{M}$.  Subscripting indices $j$ and $k$ are understood to take all values in $1, \ldots, d$.

\begin{algorithm}
\caption{Automated Blocking}\label{alg:autoblock}
\begin{algorithmic}[1]
\State $i \gets 0$
\State $\Psi_0 \gets \Psi_\text{scalar}$
\State \emph{loop}:
\State \indent $i \gets i + 1$
\State \indent Generate samples of $\Theta$ under $\Psi_{i-1}$, where $X_j$ represents the sample chain of $\theta_j$
\State \indent Discard initial 50\% of each chain $X_j$
\State \indent $\rho_{j,k} \gets \text{cor}(X_j, X_k)$
\State \indent $d_{j,k} \gets 1 - | \rho_{j,k} |$
\State \indent Construct distance matrix $D$ from elements $d_{j,k}$
\State \indent Construct hierarchical clustering tree $T$ from $D$
\State \indent $\boldsymbol{\Psi}_\text{cand} \gets \{ \Psi(T_{\text{cut}=0}), \Psi(T_{\text{cut}=0.1}), \Psi(T_{\text{cut}=0.2}), \ldots, \Psi(T_{\text{cut}=1}) \}$
\State \indent $\Psi_{i} \gets \text{argmax}_{\Psi \in \boldsymbol{\Psi}_\text{cand}} E(\Psi)$
\State \textbf{if} $E(\Psi_{i}) > E(\Psi_{i-1})$ \textbf{and} $\Psi_{i} \neq \Psi_{i-1}$ \textbf{then goto} \emph{loop}
\State $\Psi_\text{AutoBlock} \gets \Psi_i$
\State \textbf{return} $\Psi_\text{AutoBlock}$
\end{algorithmic}
\end{algorithm}

The procedure begins with the initial MCMC algorithm $\Psi_0 = \Psi_\text{scalar}$, or scalar sampling of all model parameters; lacking prior information, this is used as the starting point for gaining insight about the posterior correlation structure.  Subsequent iterations are based upon the empirical posterior correlation produced in the previous iteration, and, to a varying degree, will institute blocks for parameter groups exhibiting sufficiency high intra-group correlations.

Prior to calculating the empirical correlation terms $\rho_{j,k}$, we discard the seemingly excessive and somewhat arbitrary initial 50\% of all samples.  This should not be confused with a traditional ``burn-in," whose purpose is to ``forget" the initial state and ensure convergence to the target distribution.  Instead, discarding these initial samples allows all adaptive scalar and block samplers ample time to self-tune, and thereby achieve their theoretically optimal algorithmic efficiency.  The choice of 50\% is largely arbitrary, and excessive in most cases, and could almost certainly be relaxed without affecting algorithm performance.

Empirical correlations are transformed into distances using the transformation $d_{j,k} = 1 - | \rho_{j,k} |$.  The form of this transformation is selected to induce several properties for elements of the distance matrix $D$: the main diagonal consists of zeros; strong correlation results in $d \approx$ 0; weak or zero correlation results in $d \approx$ 1; and correlations of $\rho$ and $-\rho$ result in the same distance.

We use the R function \texttt{hclust} to create the hierarchical tree $T$ from the distance matrix $D$.  The default ``complete linkage" clustering \citep[chapter 4]{Everitt2011} is appropriate, since this ensures that all parameters within each cluster have a minimum absolute pairwise correlation.  At height $h \in [0,1]$ in $T$, the absolute correlation between parameter pairs (within clusters) is at least $1 - h$.

We use the R function \texttt{cutree} for cutting the hierarchical clustering tree $T$ at a specified height $h \in [0,1]$ to produce disjoint parameter groupings, which may be used to define parameter blocks for the purpose of MCMC sampling.  We justify this means of generating parameter sampling blocks, insofar as to increase algorithmic efficiency we strive to group \emph{correlated} parameters into sampling blocks -- the exact effect of cutting $T$ at any particular height.

We define the MCMC algorithm $\Psi(T_{\text{cut}=h}) \in \boldsymbol{\Psi}_\mathcal{M}$ as the unique MCMC algorithm defined by scalar and/or block sampling applied to the parameter blocks that result from cutting $T$ at height $h$.  We note that for all $T$, $\Psi(T_{\text{cut}=0}) = \Psi_\text{scalar}$, and $\Psi(T_{\text{cut}=1}) = \Psi_\text{block}$.

There is no universally optimal value of the cut height $h$, as our findings in section \ref{sec:efficiency} imply that any $h \in [0,1]$ may maximize the efficiency $E(\Psi(T_{\text{cut}=h}))$ for a particular model $\mathcal{M}$.  We recognize that a combination of distinct cut heights applied to different branches of $T$ may produce the maximal efficiency, but we do not consider such strategies herein.

Rather than attempting to infer what might be an appropriate cut height for model $\mathcal{M}$, we consider a range of potential cut heights, and the resulting MCMC algorithms.  These comprise $\boldsymbol{\Psi}_\text{cand}$, the candidate set of MCMC algorithms for a particular iteration.  This approach allows the blocking procedure to adjust according to the model, the posterior correlation structure, and the underlying computational architecture.  The MCMC algorithm for each iteration $(i \geq 1)$ is selected from among $\boldsymbol{\Psi}_\text{cand}$ as that which produces the maximal overall efficiency.

To estimate the integrated autocorrelation time, and hence the algorithmic efficiency, of a chain of MCMC samples, we use the \texttt{effectiveSize} function in the R \texttt{coda} package \citep{Plummer2006}.  The approach underlying this function -- using a fitted autoregressive model to estimate the spectral density at frequency zero -- has been seen to converge fastest among several methods to the true integrated autocorrelation time \citep{Thompson2010}.

As $E(\Psi_i)$ increases through the course of iterations, improved estimates of the posterior correlation are produced, giving the potential for more refined parameter blockings, and thus progressive increases in $E(\Psi_i)$ in subsequent iterations.  This iterative procedure continues until either (1) $\Psi_{i} = \Psi_{i-1}$ (identical algorithms are selected on consecutive iterations), or (2) $E(\Psi_{i}) < E(\Psi_{i-1})$ (efficiency decreases between iterations).  In practice, our procedure typically halts with terminating condition (1).  This may be concurrent with terminating condition (2), on account of stochastic variation in sampling and/or runtime.

We select the output from our automated blocking procedure as $\Psi_\text{AutoBlock}$, the MCMC algorithm selected in the final iteration.  In our experience, $\Psi_\text{AutoBlock}$ is typically identical to the MCMC algorithm of the second-to-last iteration; that is, the procedure has converged to a stationary state.  If a situation arises where the final iteration produces a different MCMC algorithm with efficiency inferior to that of the previous iteration, then prudence would suggest a thoughtful examination of the posterior samples, empirical correlation matrices, properties of the adapted samplers, convergence diagnostics, etc.

\section{Automated Blocking Performance} \label{sec:performance}

We now compare the performance of MCMC algorithms produced using the automated blocking procedure of section \ref{sec:autoblock} against various static MCMC algorithms.  First, we describe the computing environment in which our analyses are performed.  We then describe a broadly representative suite of example models, and present the performance results of automated and static MCMC algorithms for each.  A public Github repository containing scripts for reproducing our results may be found at \texttt{https://github.com/danielturek/automated-blocking-examples}.

\subsection{Computing Environment}

Since one of our points is that optimal design of MCMC algorithms depends on the computing environment, we briefly summarize the software tools and computing platform used.  All statistical models and MCMC algorithms were built using the NIMBLE package \citep{NIMBLEDevelopmentTeam2014} for R \citep{RCoreTeam2014}. NIMBLE allows hierarchical models to be defined within R using the BUGS model declaration syntax introduced by the BUGS project \citep{Lunn2000, Lunn2012}.  MCMC algorithms in NIMBLE are written using NIMBLE's domain specific language for specifying hierarchical model algorithms.  This language is an enhanced subset of R (interfaced through an R session) which is compiled into C++ code, which is subsequently compiled and run.

As a result, the examples here use highly efficient code generated automatically for each model and algorithm.  Of particular importance is that matrix operations are done via the highly optimized C++ Eigen library \citep{Guennebaud2010}.  Finally, the high-level programmability provided by R facilitated the dynamic exploration of MCMC algorithms.  Examples were run using R version 3.1.2, using the BLAS (Basic Linear Algebra Subprograms) provided by R for multivariate density calculations and simulation, and running under Macintosh OSX version 10.9.5 on a 2.5 GHz Intel Core i7 processor.

\subsection{Model Descriptions}

We tested the automated blocking procedure on seven examples, including real-data and toy models, and compared the results against standard MCMC algorithms.  When there are obvious ``hand-tuned" algorithms that a seasoned MCMC practitioner would consider for a particular model, we included those as well.  For the toy models, the goal was to construct posterior distributions with specific correlation structures as described below.  In these cases the models are simply prior distributions without any likelihood component.

\subsubsection*{Varying Size Blocks of Fixed Correlation}

This model structure demonstrates parameter groups of varying size, where each group exhibits a fixed intra-group pairwise correlation.  The model contains $N=64$ parameters, half of which are grouped to have pairwise posterior correlation of $\rho$.  This is accomplished using a prior multivariate normal distribution with appropriate covariance (equivalently, correlation) matrix, which in the absence of a likelihood term fully determines the posterior distribution for these 32 parameters.  Similarly, additional disjoint groups of correlated parameters are constructed of sizes 16, 8, 4, and 2.  The remaining two parameters are uncorrelated to any others, specified using univariate normal prior distributions.  We consider three values for the intra-group correlation, $\rho$ = 0.2, 0.5, and 0.8.  As these models have no likelihood, we are using MCMC to sample from the prior distribution.  We note that the dependence structure is the same as the block-diagonal covariance structure (with the blocks having compound symmetry) obtained when analytically integrating over the exchangeable prior means of clustered random effects.  This example thereby mimics the structure found in basic multilevel hierarchical models, albeit without the explicit computational expense of a likelihood calculation.

\subsubsection*{Fixed Size Blocks of Varying Correlation}

The next model structure exhibits fixed size groupings of parameters, with posterior correlations ranging between 0 and 0.9.  Each such model contains $N = 10n$ parameters.  Again employing multivariate normal distributions, we induce nine disjoint groupings of $n$ parameters each, having intra-group pairwise correlations of 0.1, 0.2, $\ldots$, and 0.9.  The remaining $n$ parameters are fully uncorrelated.  Three such models of this structure are constructed, using the values $n$ = 2, 5, and 10.  As in the previous models, these do not include any likelihood term.

\subsubsection*{Random Effects Model}

We select the ``litters" model from among the original example models provided with the MCMC package WinBUGS.  This random effects model contains two groups of 16 binomial observations.  Within each group, the binomial probabilities are modeled as random effects arising from a beta distribution.  The particular parameterization of the beta distributions (in terms of $\alpha$ and $\beta$, rather than $\mu$ and $\sigma$) results in strong correlations between each $\alpha_i, \beta_i$ pair.  The WinBUGS manual comments upfront that this model exhibits slow mixing.  We consider an informed MCMC algorithm, which blocks each $\alpha_i, \beta_i$ pair.  In addition, the beta-binomial conjugacy relationships permit use of cross-level sampling, where we jointly sample top-level parameters and conjugate latent states, as used by \citet[p.141-143]{Rue2005}.

\subsubsection*{Auto-Regressive Model}

We select the ``ice'' model from among the examples provided with WinBUGS as an auto-regressive \citep[AR;][]{Harvey1993} example, which is also analyzed in \citet{Breslow1993}.  The data contains 77 incident counts of breast cancer occurring in Iceland, which are modeled as Poisson counts.  Explanatory variables include age group, year of birth (represented using 11 cohorts ranging between 1840 and 1949), and the total person-years for the subjects in each group.  The model uses second-order AR smoothing of birth cohort effects.

\subsubsection*{Linear Gaussian State Space Models}

We construct two linear Gaussian state space models \citep{Durbin2012} each consisting of 100 latent states and observations.  State transitions are governed by a first order AR process, and we seek inferences about the transition process, and the system and observation noise.  We consider two equivalent parameterizations of the state transition process.  First, in terms of the intercept and mean of the AR process, which have largely uncorrelated posteriors (independent parameterization), and second, in terms of the intercept and autocorrelation, which are known to be highly correlated (correlated parameterization).  For the correlated parameterization, we consider an informed MCMC algorithm, which blocks the intercept and autocorrelation parameters.  We deliberately include this inferior parameterization, to assess MCMC performance in the case of known strong posterior correlation.  In practice, an analyst may not know which model parameterization(s) will produce uncorrelated posterior dimensions.

\subsubsection*{Spatial Model}

We consider a spatially dependent hierarchical model.  The data consist of 148 measurements of scallop abundance at various locations off the New York and New Jersey coastline, and was collected by the Northeast Fisheries Science Center of the National Marine Fisheries Service in 1993.  The data set is publicly available at \texttt{http://www.biostat.umn.edu/{\textasciitilde}brad/data/myscallops.txt}, and is analyzed in \citet{Banerjee2003}, pages 44-65.  Following \citet{Banerjee2003}, we model the mean log-abundance as multivariate normal with covariance that decays exponentially as a function of distance. The covariance is given by $\text{cov}(g_i,g_j) = \sigma^2 \exp(-d_{i,j}/\rho)$, where the observations are modeled as Poisson counts $y_i \sim \text{Poisson}(\text{exp}(g_i))$, and $d_{i,j}$ is the distance between observations $y_i$ and $y_j$.  Since this covariance structure induces a trade-off between $\sigma$ and $\rho$, we expect these parameters to be correlated in the posterior distribution.

\subsubsection*{Generalized Linear Mixed Model}

We include a reasonably sized generalized linear mixed model \citep[GLMM;][chapter 6]{Gelman2006}.  We make use of the Minnesota Health Plan dataset available in \citet{Waller1997} and follow the analysis of \citet{Zipunnikov2006}.  The dataset contains 968 counts of senior-citizen clinical visits, which are modeled as Poisson counts.  The linear predictor contains fixed and random effects, using a variety of covariates and including several interaction terms.

\subsection{Performance Results}

We present three quantities to gauge the performance of MCMC algorithm $\Psi$.  Rather than algorithmic efficiency $A(\Psi)$, for convenience of interpretation we present the proportional quantity ESS = 10,000 $A(\Psi)$, where ESS denotes effective sample size.  This scaling of $A(\Psi)$ has a natural interpretation as the number of effective samples (for the slowest mixing parameter) which result from a chain of 10,000 MCMC samples.  Similarly, to represent the computational requirement $C(\Psi)$, we present the proportional quantity Runtime = 10,000 $C(\Psi)$, interpretable as the time (in seconds) required to generate 10,000 MCMC samples.  We directly present the overall MCMC efficiency as Efficiency = ESS / Runtime = $A(\Psi) / C(\Psi) = E(\Psi)$, which is independent of any scaling, and maintains the intuitive interpretation as the number of effective samples generated per second of algorithm runtime (again, for the slowest mixing parameter).  MCMC sampling is performed using a fixed random number seed and identical initial values for each model, so identical MCMC algorithms will produce identical sample chains, and hence ESS, but not necessarily Runtime or Efficiency on account of discrepancies in algorithm runtime.  We observe the automated procedure producing the same MCMC algorithm across repeated experiments, with numerical results for Runtime and Efficiency varying less than 5\% from those presented herein.

For each example model $\mathcal{M}$, we present results for MCMC algorithm $\Psi_\text{block}$ denoted as ``All Blocked," and those of $\Psi_\text{scalar}$ as ``All Scalar," noting that $\Psi_\text{scalar}$ also represents the initial state ($0^{th}$ iteration) of the automated blocking procedure.  The maximally efficient algorithm generated via automated blocking is presented as ``Auto Blocking," which will generally represent a dynamically determined blocking scheme.  We also present a third static MCMC algorithm, which is not necessarily a member of $\boldsymbol{\Psi}_\mathcal{M}$ on account of the possible use of conjugate sampling.  This algorithm assigns block samplers to groups of parameters arising from multivariate distributions, scalar samplers to parameters arising from univariate distributions, and assigns conjugate samplers whenever the structure of $\mathcal{M}$ permits; this static algorithm may be more representative of default MCMC algorithms provided by software packages, and is denoted as ``Default."  Finally, for several example models we include an informed blocking of the model parameters, based upon expert or prior knowledge, which is referred to as ``Informed Blocking."  Results for the random effects model also include the ``Informed Cross-Level" MCMC algorithm which makes use of cross-level sampling, which is not in $\boldsymbol{\Psi}_\mathcal{M}$.

\subsubsection*{Varying Size Blocks of Fixed Correlation}

\begin{figure}[h]
\centerline{\includegraphics[scale=1.0]{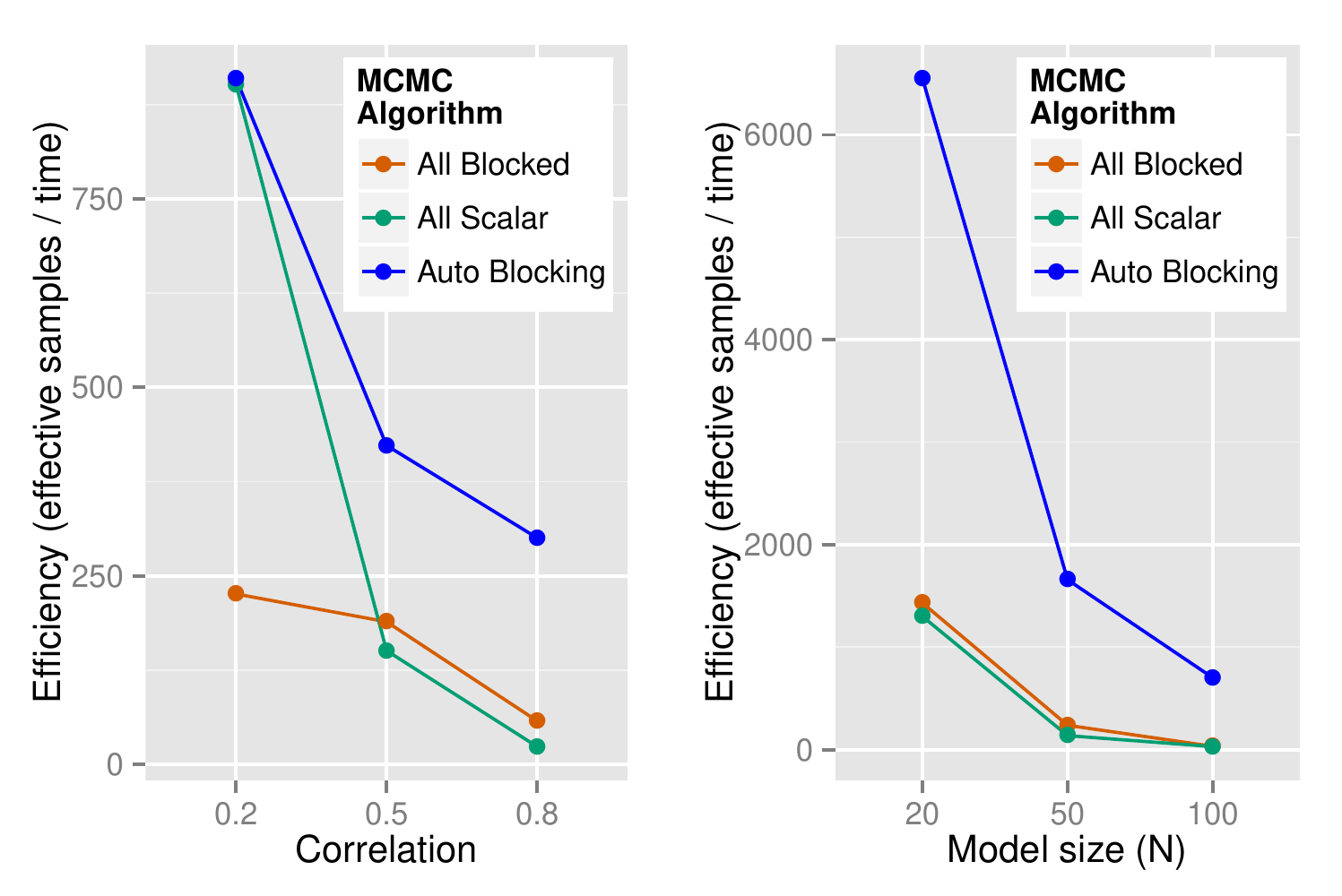}}
\caption{Efficiency results for two contrived model structures: varying sized blocks of fixed correlation (left), and fixed sized blocks of varying correlation (right).}
\label{fig:contrivedModels}
\end{figure}

The left pane of Figure \ref{fig:contrivedModels} displays the Efficiency performance for the model structures containing varying sized blocks of fixed correlation.  For $\rho=0.2$, the Auto Blocking algorithm selects cut height $h = 0$, which corresponds to re-selecting the algorithm All Scalar.  Since this MCMC algorithm is identical to the initial state, the automated procedure terminates there.  The All Blocked scheme actually runs faster, but the algorithmic efficiency loss inherent to large block sampling dominates, resulting in Efficiency approximately four times lower.  For larger values of $\rho$, the All Scalar algorithm suffers progressively more since it fails to institute any blocking in the presence of increasing correlations.  For $\rho = 0.5$ and 0.8, Auto Blocking algorithm selects cut heights $h=0.6$ and $h=0.3$, respectively, which each exactly place all correlated terms into sampling blocks.  In every case, the slowest mixing parameter is from among the largest correlated group of 32 parameters.

\subsubsection*{Fixed Size Blocks of Varying Correlation}

The right pane of Figure \ref{fig:contrivedModels} presents results for the model structure containing fixed size parameter groupings with correlations between 0 and 0.9.  For each size model, the automated blocking procedure selects a particular cut height (and hence, MCMC algorithm) twice consecutively, thus terminating on the third iteration.  The cut heights selected for models $N$ = 20, 50, and 100 are $h$ = 0.5, 0.8, and 0.9, respectively (not shown), progressively pushing more of the correlated parameter groupings into sampling blocks.  The Auto Blocking algorithm produces increases in Efficiency by factors of 4.5, 7, and 21 in the three models, over the static All Scalar and All Blocked algorithms.

\subsubsection*{Random Effects Model}

In the random effects model (Table \ref{tab:exampleModels}), automated blocking generates an MCMC algorithm identical to the Informed Blocking algorithm (blocking each $\alpha_i$, $\beta_i$ pair), which produces a tenfold improvement in Efficiency over the most efficient static algorithm -- for this model, All Scalar sampling.  The cut height $h = 0.1$ indicates that only the $\alpha_i$, $\beta_i$ pairs exhibit posterior correlations above 0.9.  The Informed Cross-Level algorithm requires a substantially longer Runtime and produces a high ESS, which results in nearly identical Efficiency as the efficiently blocked Auto Blocking algorithm.

\subsubsection*{Auto-Regressive Model}

In the auto-regressive model (Table \ref{tab:exampleModels}), an AR process value exhibited the slowest mixing under All Scalar sampling.  When all 24 model parameters (AR process values, fixed effects, and one hyper-parameter) are blocked, the algorithm Runtime is nearly halved.  This decrease in Runtime is largely due to the dependency structure inherent to the AR process.  Scalar sampling of AR process values requires nearly a three-fold increase in density evaluations of the process values (since it's a second-order AR process) relative to All Blocked sampling.  In addition to the improved Runtime, the All Blocked sampling of the correlated AR process values increases their individual algorithmic efficiencies, and the slowest mixing parameter is among the fixed effects.  The Efficiency under All Blocked sampling is over double that of All Scalar sampling.  The automated blocking procedure identifies a blocking scheme which blocks together all AR process values and fixed effects (23 total; cut height $h$ = 0.4), and performs univariate sampling of the single hyper-parameter.  This has a similar Runtime to All Blocked sampling, but increases algorithmic efficiency for all parameters.  The resulting overall Efficiency under the Auto Blocking MCMC algorithm is over three times that of All Scalar sampling.

\begin{table}[h]
\centering
\begin{tabular}{ c l r r r }
Model & MCMC Scheme & ESS & Runtime & Efficiency \\ \hline \hline
\multirow{6}{*}{\begin{tabular}[c]{@{}c@{}}Random\\Effects\end{tabular}} & All Blocked & 0.4 & 0.29 & 1.3 \\
 & Default & 1.1 & 1.19 & 1.0 \\
 & All Scalar & 2.1 & 0.51 & 4.2 \\
 & Informed Blocking & 19.0 & 0.50 & 38.2 \\
 & Informed Cross-Level & 101.3 & 2.64 & 38.5 \\
 & Auto Blocking & 19.0 & 0.48 & 39.2 \\ \hline
 \multirow{3}{*}{\begin{tabular}[c]{@{}c@{}}Auto-\\Regressive\end{tabular}} & All Blocked & 8.9 & 0.3 & 27.3 \\
 & All Scalar & 6.5 & 0.6 & 11.5 \\
 & Auto Blocking & 12.7 & 0.3 & 37.5 \\ \hline
 \multirow{4}{*}{\begin{tabular}[c]{@{}c@{}}State Space\\Independent\end{tabular}} & All Blocked & 0.3 & 0.8 & 0.4 \\
 & Default & 27.6 & 4.6 & 6.0 \\
 & All Scalar & 20.2 & 1.3 & 15.7 \\
  & Auto Blocking & 29.1 & 1.3 & 22.4 \\ \hline
\multirow{5}{*}{\begin{tabular}[c]{@{}c@{}}State Space\\Correlated\end{tabular}} & All Blocked & 0.6 & 0.7 & 0.8 \\
 & Default & 1.7 & 4.9 & 0.4 \\
 & All Scalar & 1.1 & 1.3 & 0.8 \\
 & Informed Blocking & 18.4 & 1.2 & 15.6 \\
 & Auto Blocking & 26.1 & 1.2 & 20.9 \\ \hline
\multirow{4}{*}{Spatial} & All Blocked & 0.2         &    5.71    &       0.04 \\
 & Default & 0.4         &   10.86   &       0.04 \\
 & All Scalar & 171.3     &  83.87    &      2.0 \\
 & Auto Blocking & 1208.0   &  78.62    &     15.4 \\ \hline
  \multirow{3}{*}{\begin{tabular}[c]{@{}c@{}}GLMM\end{tabular}} & All Blocked & 2.2 & 44.3 & 0.05 \\
 & All Scalar & 60.9 & 22.6 & 3.0 \\
 & Auto Blocking & 60.9 & 22.6 & 3.0 \\ \hline
\end{tabular}
\caption{MCMC performance results for the suite of example models.  Effective sample size (ESS) is measured in effective samples per 10,000 iterations, Runtime is presented as seconds per 10,000 iterations, and Efficiency is in units of effective samples produced per second of algorithm runtime.}
\label{tab:exampleModels}
\end{table}

\subsubsection*{Linear Gaussian State Space Models}

Table \ref{tab:exampleModels} presents results for both parameterizations of the state space model.  In the Independent parameterization, the observation noise parameter is the slowest mixing, in all except the All Blocked algorithm.  The All Blocked algorithm runs quickly, but is limited by the extremely low algorithmic efficiency of the AR process intercept parameter.  The Default algorithm assigns conjugate normal samplers to each latent state, resulting in high algorithmic efficiency but a substantially longer Runtime, which diminishes the overall Efficiency.  Auto Blocking (cut height $h$ = 0.8) creates a block of six parameters containing five latent states and the observation noise, and a disjoint block of the two AR process parameters.  This combination, unlikely to be discovered though any combination of prior knowledge or trial and error, produces a 40\% increase in Efficiency over All Scalar sampling, which is the most efficient static MCMC algorithm.

We suspect the intercept and autocorrelation parameters of the AR process to be correlated in the na\"ive parameterization of the state space model.  The All Blocked algorithm once again runs quickly, but is limited by the ESS of the AR process intercept.  The Default algorithm is again slow due to conjugate sampling, but similar to the All Scalar algorithm, produces low algorithmic efficiency of the correlated AR process parameters.  The Auto Blocking algorithm (cut height $h$ = 0.1) selects the same parameter block as in the Informed Blocking algorithm (AR process intercept and autocorrelation), and additionally a block containing the observation noise and a latent state.  The Runtimes are, accordingly, nearly identical, however the ESS of the observation noise, the limiting parameter, increases.  Automated blocking produces Efficiency over 20 times higher than the All Blocked algorithm, which is the most efficient static algorithm, and 25\% higher than the Informed Blocking algorithm.  It is important to note that the automated blocking procedure overcame the sampling inefficiencies introduced by this na\"ive parameterization, without requiring user intervention.

\subsubsection*{Spatial Model}

MCMC performance results for the spatial model (Table \ref{tab:exampleModels}) display several interesting trade-offs in MCMC efficiency.  The spatial model contains 148 latent parameters jointly following a multivariate normal distribution, and three top-level parameters that govern this distribution ($\mu$, $\sigma$, and $\rho$).  As there are no conjugate relationships between parameters, the sole difference between the All Blocked algorithm and the Default algorithm is the inclusion of these top-level parameters in the large sampling block.  Therefore, the five second difference in Runtime can be attributed to three fewer multivariate density evaluations (per MCMC iteration) under the All Blocked algorithm.  However, under either algorithm, the blocked sampling of latent parameters produces extremely low ESS values of 0.2 and 0.4 among the latent parameters.  The minimal ESS value increases by a factor of two when the top-level parameters are removed from the large block sampler, and thereby achieve better mixing.

The All Scalar algorithm frees all latent parameters from block sampling.  Each scalar sampler requires its own, independent, evaluation of the latent multivariate density, hence the Runtime of the All Scalar algorithm increases dramatically.  That being said, the ESS values of the slowest mixing latent parameters under the All Blocked and Default algorithms both increase to approximately 4000 (not shown). $\rho$ is the slowest mixing parameter under the All Scalar algorithm, with ESS increased from 139.6 (under the Default algorithm) to 171.3, even though it underwent scalar sampling in both cases; this is another example of the slowest mixing parameter affecting the algorithmic efficiency of other model parameters.

The automated blocking procedure selects cut height $h = 0.1$, which produces a single block containing $\rho$ and $\sigma$; this indicates an empirical posterior correlation of at least 0.9 between these parameters.  The ESS of $\rho$ increases to approximately 1500 (not shown).  A latent parameter once again produces the slowest mixing with ESS of 1208, which produces nearly a tenfold increase in Efficiency relative to the All Scalar algorithm.  The Runtime of the Auto Blocking algorithm decreases slightly compared to the All Scalar algorithm, since the single block sampler induces one fewer evaluation of the latent multivariate density.

\subsubsection*{Generalized Linear Mixed Model}

We first note that our GLMM model is by far the largest example considered, containing nearly 2000 stochastic model components (including observations); so we anticipate comparatively low MCMC Efficiencies regardless, since MCMC algorithms simply take time to carry out all model calculations.  For this model (Table \ref{tab:exampleModels}), the automated procedure converges on the All Scalar algorithm, which is the same as its initial state, and which produces overall MCMC Efficiency of about 3.  In hindsight this result may not surprise us, since the fully exchangeable nature of the random effects in this model does not induce correlations among the sampled parameters for this particular dataset.  Correspondingly, for a large number of un-correlated random effects, and in the absence of multivariate distributions, univariate sampling produces the highest Efficiency.  We also note that the All Blocked algorithm, which consists of a single block sampler of dimension 858, has Runtime approximately twice that of the All Scalar algorithm, and produces an overall Efficiency of approximately 0.05.

\begin{figure}[h]
\centerline{\includegraphics[scale=1.0]{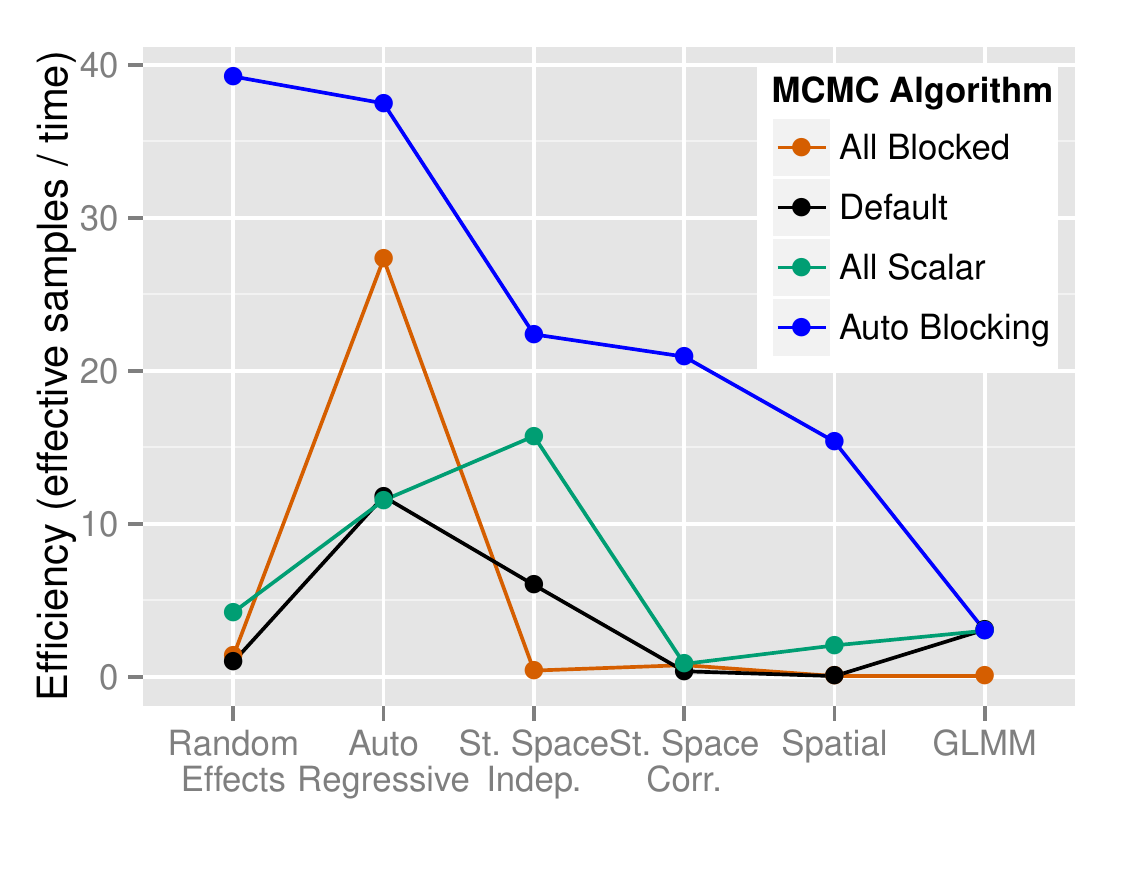}}
\caption{Efficiencies of MCMC algorithms for the suite of example models.}
\label{fig:exampleModels}
\end{figure}

\subsubsection*{Efficiency Gains from Automated Blocking}

In Figure \ref{fig:exampleModels}, we present the overall Efficiencies achieved for our suite of example models (excluding the two contrived model structures).  The Auto Blocking algorithm consistently out-performs any static algorithm in terms of Efficiency, ranging between roughly a 50\% increase to several orders of magnitude of improvement.  The exception is the GLMM example, in which Auto Blocking matches the All Scalar algorithm identically.  We observe variation in the relative Efficiencies among the static algorithms, reinforcing our notion that overall MCMC efficiency is highly dependent upon hierarchical model structure, and attempting to infer what might be an efficient MCMC algorithm for a particular problem is, in general, difficult.

\section{Discussion} \label{sec:discussion}

We have presented a general automated procedure for determining an ``efficient" MCMC algorithm for hierarchical models.  Our procedure is a greedy, iterative algorithm, which traverses a finite and well-defined set of MCMC algorithms.  This is the first such automated MCMC-generating procedure of its kind, so far as we are aware.  Using a suite of example models, we have observed that our automated procedure generates improvements in efficiency (relative to static MCMC algorithms) ranging between one and three orders of magnitude.  In each case, the automated procedure produced an MCMC algorithm at least as efficient as any model-specific MCMC algorithm making use of prior knowledge or expert opinion.  In all examples, our iterative procedure terminated within four iterations, although it is plausible that for more complex models it would proceed longer.

Our study has been confined to a single dimension of a much broader problem.  We have strictly considered combinations of scalar and blocked adaptive Metropolis-Hastings sampling, with a small number of exceptions only for the purpose of comparison (\emph{e.g.}, the use of conjugate sampling).  No less, we have restricted ourselves to non-overlapping sampling: each model parameter may only be sampled by a single MCMC sampler function.  We may instead view the domain of our problem (automated determination of an efficient MCMC algorithm) as a broader space of MCMC algorithms.  This space may permit a wide range of sampling algorithms not considered herein: auxiliary variable algorithms such as slice sampling \citep{Neal2003}, or derivative-based sampling algorithms such as Hamiltonian Monte Carlo \citep{Duane1987}, among many possibilities.  The resulting combinatorial explosion in the space of MCMC algorithms makes any process of trial-and-error, or an attempt at comprehensive exploration, futile.  It is for this reason we seek to develop an automated procedure for determining an efficient MCMC algorithm, which may not be globally, maximally efficient, but provides non-trivial improvements in efficiency, nonetheless.

It should be noted that aspects of the problem addressed herein superficially resemble, but are fundamentally different in nature from hierarchical clustering, or sparse covariance estimation.  Granted, our automated procedure firstly utilizes an empirical covariance matrix generated from MCMC sampling chains.  However, whereas sparse covariance estimation seeks to estimate the non-zero elements of the underlying covariance structure \citep{Cai2011}, our procedure concerns the non-trivial (correlated) elements, with little concern for the smaller entries.  Our blocking algorithm also makes use of the complete linkage clustering algorithm, for determining groupings of correlated model parameters.  Clustering algorithms have been applied to a wide variety of problems \citep{Xu2005}, but not to parameters of hierarchical models specifically with the aim of accounting for trade-offs between MCMC algorithmic efficiency and computational requirements, to produce a \emph{computationally efficient} MCMC algorithm.  This is a fundamentally different goal than merely producing groupings of ``similar" parameters (given some measure of similarity), as is generally the goal in most clustering applications.  Thereby, the existing literature on these subjects is related, but not intimately applicable to our problem at hand.  A deeper consideration of these topics may be worthwhile, but we consider it beyond the scope of this paper.

Reasonably straightforward improvements could be made to our automated blocking procedure, which is presented as a sensible first approach that addresses the factors affecting MCMC algorithm efficiency.  By design, our procedure natively accounts for differences in system platform or architecture that may affect the relative efficiencies of MCMC algorithms.  We can envision a wide variety of possible extensions to our algorithm, ranging from only re-blocking the slowest mixing parameter on each iteration, to permitting cuts at different heights on distinct branches of the hierarchical clustering tree.  Our procedure is intended as a proof-of-concept for the automated generation of efficient MCMC algorithms, and to serve as a starting point for subsequent research.


\if0\blind{
\section*{Acknowledgements}
This work was supported by the NSF under grant DBI-1147230.
} \fi

\newpage
\printbibliography

\end{document}